\documentstyle[%preprint,
               twocolumn,
               epsfig,amssymb,aps]{revtex}
 
\newcommand{\etal}{{\em et al.}}
\newcommand{\dt}{\langle{t_{\rm{BaF}}}\rangle-\langle{t_\gamma}\rangle}
 
\begin{document}
 
\preprint{}\draft
 
\title{Radiative proton capture on $^6$He}
 
\author{E.~Sauvan$^a$\thanks{Present address: ISOLDE, CERN, Switzerland.},
F.M.~Marqu\'es$^a$\thanks{e-mail: {\tt Marques@caelav.in2p3.fr}},
H.W.~Wilschut$^b$, N.A.~Orr$^a$, J.C.~Ang\'elique$^a$, C.~Borcea$^c$,
W.N.~Catford$^d$, N.M.~Clarke$^e$, P.~Descouvemont$^f$, J.~D\'{\i}az$^g$,
S.~Gr\'evy$^a$, A.~Kugler$^h$, V.~Kravchuk$^b$,
M.~Labiche$^a$\thanks{Present address: University of Paisley, Scotland.},
C.~Le~Brun$^a$\thanks{Present address: ISN, Grenoble, France.},
E.~Lienard$^a$, H.~L\"ohner$^b$, W.~Mittig$^i$, R.W.~Ostendorf$^b$,
S.~Pietri$^a$, P.~Roussel-Chomaz$^i$, M.G.~Saint~Laurent$^i$, H.~Savajols$^i$,
V.~Wagner$^h$, N.~Yahlali$^g$}
 
\address{$^a$ Laboratoire de Physique Corpusculaire,
 IN2P3-CNRS, ISMRA et Universit\'e de Caen, F-14050 Caen cedex, France}
\address{$^b$ Kernfysich Versneller Instituut, Zernikelaan 25,
 NL-9747 AA Groningen, The Netherlands}
\address{$^c$ IFIN-HH, P.O.~Box MG-6, 76900 Bucharest-Magurele, Romania}
\address{$^d$ Department of Physics, University of Surrey,
 Guildford, Surrey, GU2~7XH, U.K.}
\address{$^e$ School of Physics and Astronomy, University of Birmingham,
 Birmingham B15 2TT, U.K.}
\address{$^f$ Universit\'e Libre de Bruxelles, CP 229,
 B-1050 Bruxelles, Belgium}
\address{$^g$ Instituto de F\'{\i}sica Corpuscular, E-46100 Burjassot, Spain}
\address{$^h$ Nuclear Physics Institute, 25068 \v{R}e\v{z} u Prahy,
 Czech Republic}
\address{$^i$ GANIL, CEA/DSM-CNRS/IN2P3, BP 55027, F-14076 Caen cedex, France}
 
\date{\today}
 
\maketitle
 
\begin{abstract}
Radiative capture of protons is investigated as a probe of clustering in nuclei
far from stability. The first such measurement on a halo nucleus is reported
here for the reaction $^6$He(p,$\gamma$) at 40~MeV. Capture into $^7$Li is
observed as the strongest channel. In addition, events have been recorded that
may be described by quasi-free capture on a halo neutron, the $\alpha$ core and
$^5$He. The possibility of describing such events by capture into the continuum
of $^7$Li is also discussed.
\end{abstract}
 
 
%%%%%%%%%%%%%% Introduction:
In the vicinity of the neutron drip-line, the weak binding of valence neutrons
may lead to the formation of spatially extended nuclei \cite{Han95}. The most
exotic of these are the core-n-n halo systems, $^6$He, $^{11}$Li and $^{14}$Be,
which exhibit Borromean characteristics whereby the two-body subsystems are
unbound \cite{Zhu93}. Owing to the large cross sections, of the order of barns,
dissociation reactions have been the most widely exploited method to study the
internal correlations \cite{Sac93,Zin97,Aum99}. The task, however, in such an
approach is complicated by the interplay of the reaction mechanism and
final-state interactions (FSI) with the intrinsic structure \cite{FMM00}. The
possibility of using the interference between 2n transfer and elastic
scattering with $^6$He has also been considered \cite{Oga99,Kro00}.
%; coupled channel studies suggest, however, that the sensitivity to the
%details of the wave function is small \cite{Kro00}.
 
Recently, an investigation of coherent bremsstrahlung production in the
reaction $\alpha$(p,$\gamma$) at 50~MeV has demonstrated that the high-energy
photon spectrum is dominated by capture to form $^5$Li \cite{Hoe00}. Such
results have motivated the extension of this technique to study $^6$He. Given a
proton wavelength of $\lambdabar=0.7$~fm at 40~MeV, it may be possible to
observe direct capture, as a quasi-free process, on the constituents of $^6$He
in addition to capture into $^7$Li. Moreover, the different quasi-free capture
(QFC) processes would lead to different $E_\gamma$ in the range 20--40~MeV. In
this Letter the first experimental results for capture on a halo nucleus are
reported. Evidence for QFC on $^{4,5}$He and n is presented; capture, however,
on a di-neutron does not appear to occur. These observations suggest that
radiative capture may provide a new probe for the study of clustering in the
ground state (g.s.) of nuclei far from stability.
 
%%%%%%%%%%%%%% Setup:
The $^6$He beam (5$\times$10$^5$~pps, $\Delta{E}/E\sim1\%$) was produced by 
fragmentation of a $^{13}$C primary beam using the GANIL coupled cyclotron 
facility, and bombarded a solid Hydrogen target \cite{Lib97} with a thickness 
of 95~mg/cm$^2$; the mean beam energy at the target midpoint was 40~MeV/N. The 
different charged reaction products emitted in the forward direction 
($\pm2^\circ$) were identified, and momentum analysed, using the SPEG
spectrometer \cite{Bia89}, which covered a rigidity range of 1.45--1.85~Tm in
three overlapping settings. The photons were detected in coincidence with the
charged fragments using the ``Ch\^ateau de Cristal'' array, with the 74 BaF$_2$
crystals placed around the target at a distance of 30~cm in two domes
\cite{ChdCr}, the total efficiency being close to 70\%. The energy calibration,
in the range 1--100~MeV, was determined from the energy deposited by cosmic-ray
muons \cite{Sau00} and the 4.43~MeV $\gamma$-rays from an Am-Be source. The
energy and angle of the photons were reconstructed using a clustering algorithm
\cite{Sau00,FMM95}, with average energy and angular resolutions of 17\% and
$10^\circ$, respectively. The event trigger required an energy deposition of at
least 3~MeV in one crystal in coincidence with a fragment in SPEG. Owing to the
compact geometry of the Ch\^ateau and the energy spread in the beam
($\Delta{t}\sim6$~ns at the target position), event-by-event $\gamma$-n
discrimination was not possible. However, for each class of events the mean
flight time between the BaF$_2$ crystals and the accelerator RF signal
$\langle{t_{\rm{BaF}}}\rangle$ could be determined. A Monte-Carlo simulation
was developed, in which the response of the Ch\^ateau and the conversion of
photons in the target frame were simulated using GEANT \cite{GEANT}; the
characteristics of the secondary beam and the spectrometer acceptances were
also included \cite{Sau00}. 
 
\begin{figure}[tb]
 \begin{center}
  \mbox{\psfig{file=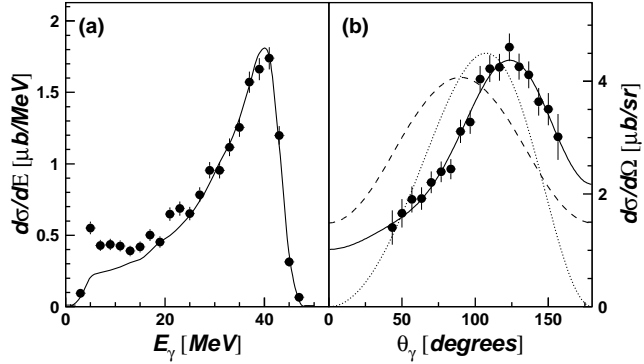,width=8.5cm}}
 \end{center}
 \caption{Energy (a) and angular distributions (b) in the $^6$He+p c.m.\ for
photons in coincidence with $^7$Li. The solid line in (a) is the response of
the Ch\^ateau to $E_\gamma=42$~MeV. The lines in (b) correspond to a classical
electrodynamics calculation (dotted), a microscopic cluster model (dashed),
both normalized to the data, and to a Legendre polynomial fit (solid).}
 \label{f:7Li}
\end{figure}
 
%%%%%%%%%%%%%%%%%%% 7Li+gamma
Turning to the experimental observations, the capture reaction
$^6$He(p,$\gamma$)$^7$Li is unambiguously identified by the $\gamma$-rays in
coincidence with $^7$Li (Fig.~\ref{f:7Li}). In particular, the photon energy
spectrum, as well as the $^7$Li momentum \cite{Sau00}, is well described
assuming a $\gamma$-ray line at 42~MeV and is free of background at higher
energies. The two particle-stable states of $^7$Li, the g.s.\ and the first
excited state at 0.48~MeV, were too close together in energy to be
distinguished in this experiment. In addition, the lowest threshold on any of
the BaF$_2$ crystals, 1~MeV, prevented observation of the 0.48~MeV
$\gamma$-ray. The photon angular distribution (Fig.~\ref{f:7Li}b) is slightly
backward peaked, as expected from classical electrodynamics \cite{Hoe99} due to
the charge asymmetry of the entrance channel. A fit with a Legendre polynomial
\cite{Wel82} leads to $a_{i=1,4}=-0.57\pm0.10$, $-0.28\pm0.13$, $0.37\pm0.10$,
$-0.21\pm0.12$, respectively. The total efficiency for the detection of
$^7$Li-$\gamma$ coincidences was estimated to be $37\pm2~\%$, and the deduced
cross section was $\sigma=35\pm2~\mu$b. As only photons are liberated in this
reaction, the measured time spectrum served as a reference,
$\langle{t_\gamma}\rangle$, for other channels in which neutrons were also
emitted.
 
The $^6$He(p,$\gamma$)$^7$Li cross section has been calculated using a
microscopic cluster model \cite{Des95}. The $^7$Li and $^6$He+p wave functions
were defined by antisymmetric products of cluster wave functions, including
$^6$He+p, $^6$Li+n and $\alpha$+t structures. Excited states of $^6$He and
$^6$Li were included in the basis. The Minnesota interaction was used to
specify the N-N force \cite{Tho77}, with an exchange parameter $u=0.935$ and a
zero-range spin-orbit force with amplitude $S_0=38$~MeVfm$^5$. This model
provides a good description of the cross sections at low energy for
t($\alpha$,$\gamma$)$^7$Li, $^6$Li(p,$\gamma$)$^7$Be and
$^6$Li(p,$\alpha$)$^3$He. At 40~MeV, a cross section for $^6$He(p,$\gamma)^7$Li
of $\sigma=59~\mu$b is calculated, with 15~$\mu$b going to the g.s.\ and
44~$\mu$b to the first excited state. At high energy, the microscopic model is
expected to provide an upper limit to the cross section as some open channels,
such as three-body ones, are not included. The relative population
$\sigma_{0.48}/\sigma_{\rm{g.s.}}=2.9$ should, however, be more reliable. The
calculation was restricted to the dominant E1 multipolarity, thus leading to an
angular distribution symmetric about $90^\circ$ (Fig.~\ref{f:7Li}b). The cross
section to the g.s.\ can be obtained from photodisintegration \cite{Sen85} via
detailed balance considerations and is $9.6\pm0.4~\mu$b. Given the predicted
relative populations of the ground and first excited state, a total capture
cross section of $\sigma\sim38~\mu$b is obtained, in agreement with the value
measured here.
 
\begin{figure}[tb]
 \begin{center}
  \mbox{\psfig{file=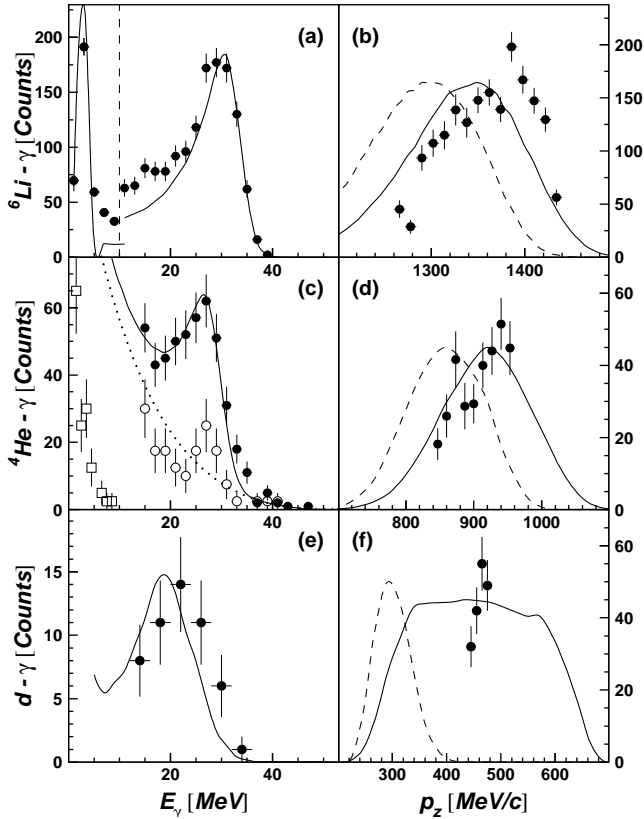,width=8.5cm}}
 \end{center}
 \caption{$\gamma$-ray energy spectrum in the $^6$He+p c.m.\ and momentum
distribution of the coincident fragment for $^6$Li (upper), $\alpha$ particles
(middle) and deuterons (lower panels). The lines correspond to calculations of
QFC on the $^5$He cluster, the $\alpha$ core and one halo neutron,
respectively, on the right with/without (solid/dashed) fragment FSI (see text).
The distribution in (a) was divided by 3 below 10~MeV, and the open symbols in
(c) are from an analysis investigating the role of the neutron background (see
text).} \label{f:QFC}
\end{figure}

%%%%%%%%%%%%%%%%%%% 6Li+gamma
QFC has been investigated by searching for $\gamma$-rays in coincidence with
fragments lighter than $^7$Li. The corresponding energy spectra
(Fig.~\ref{f:QFC}a,c,e) do indeed exhibit peaks below 42~MeV. In order to
establish the origin of these fragment-$\gamma$ coincidences, QFC processes on
the subsystems of $^6$He have been modelled as follows. The $^6$He projectile
is considered as a cluster ($A$) plus spectator ($a$) system in which each
component has an intrinsic momentum distribution, the corresponding energy
$E_A+E_a-m_{^6\rm{He}}$ being taken into account in the total available energy.
The reaction may be denoted as $a$+$A$(p,$\gamma$)$B$+$a$, and the $\gamma$-ray
angular distribution is assumed to be that given by the charge asymmetry of the
entrance channel ($A$+p) \cite{Hoe99}. The intrinsic momentum distribution of
all the clusters was taken to be Gaussian in form with ${\rm{FWHM}}=80$~MeV/$c$
\cite{Sau00}; the resolution in the measured photon energy is such that the
results are relatively insensitive to the exact value.
 
In order to explore the possibility that FSI may occur in the exit channel
between the spectator, $a$, and the capture fragment, $B$, an extended version
of the QFC calculation was developed. Here the energy available in the system
$B$+$a$ is treated as an excitation in the continuum of $^7$Li, which is
allowed to decay in flight.
 
In the case of $^6$Li-$\gamma$ coincidences, two lines were observed
(Fig.~\ref{f:QFC}a) at 30 and 3.5~MeV. These are clearly associated with the
formation of $^6$Li and the decay of the second excited state, at 3.56~MeV
\cite{Ajz88}.
%The mean flight time for events registered in the
%Ch\^ateau, $\dt=0.4\pm0.1$~ns, confirmed the detection of photons.
Taking into account the detection efficiencies, we find that the $^6$Li is
formed almost exclusively ($96^{+4}_{-24}\%$) in the 3.56~MeV excited state.
%The $^6$Li formed by (p,n) reactions was well separated in SPEG \cite{Sau00}.
The estimated cross section was $\sigma=3.5\pm1.3~\mu$b. The lines in
Fig.~\ref{f:QFC}a,b represent the results of QFC on $^5$He into
$^6$Li$^*$(3.56~MeV). The $\gamma$-ray energy spectrum is well described, while
reproducing the $^6$Li momentum distribution requires inclusion of $^6$Li-n
FSI as described above. The QFC with fragment FSI approach is thus the one
employed in the following discussions.
 
\begin{figure}[tb]
 \begin{center}
  \mbox{\psfig{file=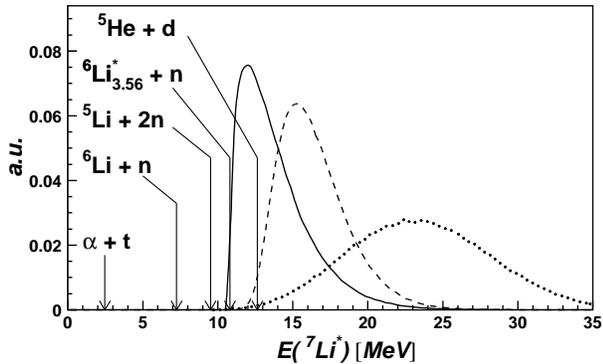,width=8.5cm}}
 \end{center}
 \caption{Relative energies within the QFC with FSI model (see text) of the
fragments in the exit channel following capture on $^5$He, the $\alpha$ core
and halo neutrons (solid, dashed and dotted lines, respectively). The various
decay thresholds in $^7$Li are indicated by the arrows.} \label{f:Eex}
\end{figure}

The apparently exclusive population of the $^6$Li$^*$(3.56 MeV) indicates the
importance of this state as the $T=1$ analogue of $^6$He$_{\rm{g.s.}}$. It has
been shown that the configuration of the 3.56~MeV state is likely to be ``a
spatially extended halolike structure formed by the neutron and proton outside
the $\alpha$ particle'' \cite{Ara95}, possibly even more extended than $^6$He.
In terms of the QFC process the population of this state is greatly favoured
owing to the overlap of the initial and final wave functions. In the case of
capture on $^6$He into the $^7$Li continuum, the reaction can proceed via the
$T=3/2$ state at 11.24~MeV \cite{Ajz88} (the $T_>$ state of
$^7$Li$_{\rm{g.s.}}$), which can only decay by neutron emission to a $T=1$
state in $^6$Li. Note, however, that the 11.24~MeV state in $^7$Li can also be
formed in the exit channel following QFC on $^5$He, as the distribution of
$^6$Li-n relative energy (Fig.~\ref{f:Eex}) is centered at $\sim12$~MeV.
 
%%%%%%%%%%%%%%%%%%% 4He+gamma
We have also searched for evidence of QFC on the $\alpha$ core, whereby the two
halo neutrons would behave as spectators. The photon spectrum should resemble
that observed for the $\alpha$+p reaction \cite{Hoe00}. Indeed such a
$\gamma$-ray energy spectrum (Fig.~\ref{f:QFC}c) has been observed in
coincidence with $\alpha$ particles. The background, however, arising from
$^6$He breakup, in which the $\alpha$ particle is detected in SPEG and the halo
neutrons interact with the forward-angle detectors of the Ch\^ateau, is
significant. In order to minimise this background, only the backward-angle
detectors ($\theta>110^\circ$) of the Ch\^ateau have been used in the analysis.
The $\gamma$-ray spectrum under this condition exhibits two components: a peak
at $E_\gamma=27$ MeV and a $1/E_\gamma$ continuum similar to coherent
$\alpha$+p bremsstrahlung \cite{Hoe00}.
 
Simulations indicate, however, that some back-scattered neutrons remain from
breakup ($\dt=0.8\pm0.1$~ns), which would also lead to a continuous component
with a $1/E$ type spectrum in the Ch\^ateau \cite{Sau00}. This would explain
why the peak-to-continuum ratio is smaller here than that found in
Ref.~\cite{Hoe00}. Therefore, we have added a single background component with
a $1/E$ form (dotted line in Fig.~\ref{f:QFC}c) to the QFC process
$\alpha$(p,$\gamma$)$^5$Li. The photon energy spectrum is thus well described,
as is the momentum distribution of the $\alpha$ particle. The cross section is
estimated to be $\sigma=4\pm1~\mu$b. Additional support may be found in
$\alpha$-$\gamma$-n coincidences, for which some 30 events are observed. Here,
the neutron was associated with events in the forward-angle detectors
($\theta<60^\circ$, $\dt=5\pm4$~ns) and the photon with events observed in
coincidence at backward angles ($\theta>110^\circ$, $\dt=1.6\pm1.4$~ns). The
resulting spectra exhibit the $1/E$ form for neutrons (open squares) and, more
importantly, a higher peak-to-continuum signal at 27~MeV for the photons (open
circles) which is closer to that measured previously for the
$\alpha$(p,$\gamma$)$^5$Li reaction \cite{Hoe00}.
 
%%%%%%%%%%%%%%%%%%% d,t+gamma
Finally, d-$\gamma$ coincidences presenting a peak in the $\gamma$-ray energy
spectrum, at $E_\gamma=20$--22~MeV, were also observed (Fig.~\ref{f:QFC}e). For
this channel the analysis was also restricted to the backward-angle detectors
($\theta>110^\circ$) of the Ch\^ateau. The relatively low statistics arise from
the limited acceptances of the spectrometer for deuterons (Fig.~\ref{f:QFC}f).
We have verified with an empty-target run that no background events are present
in the energy range in question, and that the events observed correspond to
photons ($\dt=0.0\pm0.3$~ns). The predictions for n(p,$\gamma$)d QFC on one
halo neutron present a peak at 19~MeV (Fig.~\ref{f:QFC}f). The small shift may
be attributable to the strong kinematic correlation between the deuteron
momentum and the photon energy, as the detection of a very small fraction of
the deuterons (depending on the neutron momentum distribution used) is
predicted \cite{Sau00}. Given these uncertainties, no reliable estimate of the
cross section for this channel was possible.
 
There are additional QFC channels, 2n(p,$\gamma$)t and t(p,$\gamma$)$\alpha$, 
that could have been observed with finite efficiency in this experiment but 
were not \cite{Sau00}. Perhaps the most interesting is QFC on the two halo
neutrons. In the case of $^6$He, several theoretical models predict the
coexistence of two configurations in the g.s.\ wave function: the so-called
``di-neutron'' and ``cigar'' configurations \cite{Zhu93}. Here one might expect
that the different admixtures of these could be probed by the relative strength
of the n,2n(p,$\gamma$)d,t QFC processes, whereby the corresponding free cross
sections at 40~MeV, obtained from detailed balance considerations, are
comparable: 9.6~$\mu$b \cite{Ahr74} and 9.8~$\mu$b \cite{Fau80}, respectively.
However, events registered in the Ch\^ateau in coincidence with tritons in SPEG
have energies below 10~MeV, whereas the 2n(p,$\gamma$)t reaction should produce
photons with $E_\gamma\approx32$~MeV. In addition, the flight time
$\dt=4.7\pm0.2$~ns clearly corresponded to neutrons.
 
We have seen that the QFC with fragment FSI model describes well the
monoenergetic $\gamma$-rays observed, as well as the momentum distribution of
the capture fragment ($B$). The $\gamma$-ray lines are associated with
{\em{specific}\/} energy distributions for the fragments in the exit channel
(Fig.~\ref{f:Eex}), depending on the intrinsic momenta of the clusters.
Therefore, such a process will exhibit the same kinematics as capture into
continuum states above the corresponding threshold,
$^6$He(p,$\gamma$)$^7$Li$^*$$\rightarrow$$B$+$a$, provided that the equivalent
region of the continuum (Fig.~\ref{f:Eex}) is populated. If, however, all the
final states observed here were the result of radiative capture into $^7$Li,
capture via the non-resonant continuum in $^7$Li might well be expected to
occur \cite{Sid86}. This would lead to a continuous component to the
$\gamma$-ray energy spectra. Moreover, events corresponding to
$E_{^7\rm{Li}^*}=0.5$--10~MeV have not been observed in either t-$\gamma$
coincidences or $\alpha$-$\gamma$ coincidences with $E_\gamma=32$--42~MeV, nor
has the decay into $\alpha$+t for $E_{^7\rm{Li}^*}>10$~MeV. Within the picture
of QFC on clusters, this is simply explained by the absence of the
2n(p,$\gamma$)t and t(p,$\gamma$)$\alpha$ QFC processes for the $^4$He-2n
\cite{Zhu93} and t-t \cite{Ara99} configurations, respectively, indicating that
$^4$He-n-n is the dominant configuration in $^6$He. This in agreement with the
relatively large n-n distance found in Ref.~\cite{FMM00}.
 
%%%%%%%%%%%%%%%%%%% conclusions
In summary, radiative capture of protons on a halo nucleus, $^6$He, has been
measured for the first time. In addition to the $^6$He(p,$\gamma$)$^7$Li
reaction, evidence for QFC on subsystems ($^5$He, $\alpha$ and n) of $^6$He has
been found. Of particular importance is the observation of events which
correspond to the previously measured $\alpha$(p,$\gamma$) reaction, as well as
the non-observation of capture on a di-neutron. Theoretically, microscopic
models need to be developed in order to describe capture on the constituent
clusters of exotic nuclei and, for comparison, capture on the projectile into
unbound final states. In this context, better knowledge of the high-lying
continuum of $^7$Li would be very helpful. Finally, it would be of interest to
study the evolution with beam energy of the contribution of QFC on clusters.
 
The support provided by the technical and operations staff of GANIL and
LPC-Caen is gratefully acknowledged. Additional support from the Human Capital
and Mobility Programme of the European Community (contract n$^\circ$
CHGE-CT94-0056) and the GDR Noyaux Exotiques (CNRS-CEA) is also acknowledged.
 
%%%%%%%%%%%%%%%%%%%%%%% biblio %%%%%%%%%%%%%%%%%%%%%%%%%%%%%%%%%%%%%%%%%%%

\end{document}